\documentclass[singlecolumn,nofootinbib]{revtex4-2}  

\usepackage{graphicx}
\usepackage[english]{babel}
\usepackage[utf8]{inputenc}
\usepackage[colorinlistoftodos, color=green!40, prependcaption]{todonotes}
\usepackage{amsmath}

\usepackage{amsfonts}
\usepackage{amssymb}
\usepackage{dsfont}
\makeatletter\AtBeginDocument{\let\@elt\relax}\makeatother
\usepackage{bbold}
\usepackage{dcolumn}
\usepackage{bm}
\usepackage[colorlinks=true,linkcolor=blue,urlcolor=blue,citecolor=blue]{hyperref}
\usepackage[normalem]{ulem}

\begin{document}
\title{The universe is not a Lie,\\
	but actually an Hopf, algebra}

\author{Niccol\`o Loret}
\email{niccolo.loret@iss.it}
\affiliation{Natl. Center for Radiation Protection and Computational Physics, Istituto Superiore di Sanità (ISS), Italian Institute of Health, Viale Regina Elena 299, 00161 Rome, Italy}

\begin{abstract}
	\noindent
	In this paper I would like to show how the Deformed Special Relativity family of models - developed to approach spacetime quantization - can actually be applied to the description of classical cosmology.
	I use the bicrossproduct basis of $\kappa$-Poincar\'{e} algebra to describe photon propagation in deSitter classical General Relativity. I show the Hopf algebraic aspects of deSitter model, and give an explicit physical interpretation of $\kappa$-Poincar\'{e} Hopf algebraic features in spacetime. Such an approach allows to unravel some not yet known General Relativistic relations of deSitter cosmology, and reinterpret many features of Quantum Gravity phenomenology as classical properties of maximally symmetric spacetime models. In the last section of the paper I give a first example on how to apply this mathematical framework to more realistic (non maximally symmetric) spacetime models, such as $\Lambda\text{CDM}$ and matter-dominated universe.
\end{abstract}

\maketitle
\tableofcontents

\newpage

\section{Introduction}
\noindent 
First of all, this paper is not about Quantum Gravity. This paper is about applying a well defined mathematical model, $\kappa$-Poincar\'{e} Hopf algebra \cite{LukNowRueggTol,Lukierski:1992dt,Luk1993,Luk1994}, very well known and studied in the Quantum Gravity literature, to the study of the signal-exchange between observers in Cosmology and related relativistic features.\\
$\kappa$-Poincar\'{e} is a deformation of Poincar\'{e} algebra in which momentum-space has a deSitter-like geometry \cite{LukDSR,KowalskiDSR}. This mathematical model can be considered as the backbone of the Deformed Special Relativity (DSR) approach to Quantum Gravity \cite{KowalskiDSR,LukDSR,AmelinoDSR2001,AmelinoDSR2002,SmMagDSR2001,SmMagDSR2003}.
More or less a decade ago, with my former research group, we showed \cite{Lateshift} how one of the most studied phenomenological features of DSR, the relativistic signals energy-dependence ({\it see} for instance \cite{AmelinoDawn,AmelinoDSR2010,Hossenfelder,Bob,kBob}), was nothing but the dual effect in momentum-space than deSitter cosmological redshift. We showed how just exchanging the cosmological constant $H_\Lambda$ with the DSR deformation parameter $\ell$, energy with time, and momenta with particles' position, the mathematical description of the two effects is basically the same. This interpretation provided us a deeper understanding of DSR approach to Quantum Gravity, since in this framework, the foundation of DSR, the highest achievable energy scale (often identified with Planck energy $E_P = \sqrt{\hslash\cdot c^5/G}$), becomes in turn nothing but the momentum-space event horizon.\\
At this point one may ask: {\it Does it mean that $\kappa$-Poincar\'{e} is just deSitter mathematical model in momentum-space?} The most common and simplistic answer to this question usually is that the resemblance between the two models is just apparent since deSitter one is a Lie algebra, while $\kappa$-Poincar\'{e} is in the matter of facts a more complex Hopf algebra \cite{Fuchs,MajidQGP}.\\
Are we sure about that? I am not questioning $\kappa$-Poincar\'{e} being an Hopf algebra, of course it is, but are we so sure that beneath deSitter model it is not possible to unravel any Hopf algebraic aspect? The aim of this paper is precisely to show how also Hopf algebra features like coproduct, antipode and counity can find a compelling interpretation in the framework of curved spacetime relativity. Such an analysis on the one hand is useful to better interpret phenomenology tricky QG phenomenology effects - such as particles vertices in DSR - by comparison with with more familiar phenomena in deSitter; on the other hand Hopf algebraic features could provide a wide set of tools for the study of cosmology and General Relativity.\\
With respect to our (already cited) previous analysis \cite{Lateshift}, an important further element to be introduced here is the duality between particles in $\kappa$-Poincar\'{e} and observers in deSitter.\\

This paper is composed by an introductory part: sections II, III; a more theoretical exploration: sections IV, V, VI and VII; and a more phenomenological technical application of DSR features in cosmology in section VIII. In the first two parts I will adopt a set of coordinates in which $c=\hslash=1$, only in Sec. VIII I will switch back to the International System of Units, in order to obtain worldlines in terms of Mpc/year.\\
In section II a short review of deSitter model is sketched, introducing a few useful elements for the following analysis. In section III are instead briefly reviewed all the fundamental aspects of $\kappa$-Poincar\'{e} Hopf algebra, as well as the basic result obtained in our previous analysis \cite{Lateshift}, also called the lateshift effect (time delay as redshift in momentum-space). In section IV is analyzed the deformed composition law due to non trivial coproduct. Section V and VI explore a not-well-known (and pretty complex) relativistic feature such as backreaction, at first in $\kappa$-Poincar\'{e} and then in deSitter. Section VII summarizes the coalgebric aspects of deSitter spacetime and goes into detail on the parallelism between particles in DSR and observers in deSitter. In conclusion in section VIII is addressed the problem of how to apply those theoretical speculations to the study of signal propagation in a phenomenological framework in which astrophisical data are only characterized by measured distance and redshift, and in section IX is shown a first attempt to apply Hopf algebraic features to more realistic universe models such as matter-dominated universe and $\Lambda\text{CDM}$.

\section{deSitter spacetime}
\noindent 
In order to fix the notation, we resume here the FLRW equations for a homogeneous and isotropic universe can be obtained solving Einstein equation
\begin{equation}
R_{\mu\nu}-\frac{1}{2}g_{\mu\nu}R-\Lambda g_{\mu\nu}=\chi T_{\mu\nu}\,,
\end{equation}
where $\Lambda$ is the cosmological constant, looking for a general solution for the metric, formalized by the line element
\begin{equation}
ds^2=(dx^0)^2-a(x^0)^2\left(\frac{dr^2}{1-k r^2}+r^2 d\Omega^2\right)\,,
\end{equation}
where $a(x^0)$ is the universe radius of curvature (or scale factor), $k$ is the Gaussian curvature which can be (-1,0,1) for negative, zero or positive curvature and, in order to satisfy the homogeneity and isotropy requirements, one can choose a perfect-fluid like stress-energy tensor:
\begin{equation}
T_{\mu\nu}=\left(\begin{array}{cccc}
\rho & 0 & 0 & 0\\
0 & -\frac{P}{1 - k r^2} a^2  & 0 & 0\\
0 & 0 & -P a^2 r^2  & 0\\
0 & 0 & 0 & -P a^2 r^2\sin^2{\theta} 
\end{array}\right)\,,
\end{equation}
in which, $\chi=8\pi G/c^4$ and $P,\rho$ are the fluid's energy density and pressure.
Einstein equation then reduces to the two FLRW equations:
\begin{eqnarray}
&&\left(\frac{\dot{a}}{a}\right)^2+\frac{k}{a^2}=\frac{\chi}{3}\rho+\frac{\Lambda}{3}\,,\label{modFried1}\\
&&2 \frac{\ddot{a}}{a}+\frac{k}{a^2} + \frac{\dot{a}^2}{a^2} = \Lambda -\chi P\,.\label{modFried2x}
\end{eqnarray}
Equations (\ref{modFried1}) and (\ref{modFried2x}) can easily be coupled, obtaining
\begin{equation}
\frac{\ddot{a}}{a} = \frac{\Lambda}{3}-\frac{\chi}{6}(\rho + 3 P)\,,
\end{equation}
which in turn, coupled with the first law of thermodynamics for an adiabatic expansion of the universe gives:
\begin{equation}
\dot{\rho}=-3\frac{\dot{a}}{a}\left(\rho + P\right)\,.
\end{equation}
The flat spacetime solution ($k = 0$) in absence of matter ($\rho,P=0$) for the above equations implies an exponential scale factor
\begin{equation}
a(x^0)=e^{\pm \sqrt{\frac{\Lambda}{3}}x^0}\,,\label{scfactds}
\end{equation}
which specializes the family of Friedmann spacetimes to the maximally symmetric de Sitter spacetime (see for an interesting insight \cite{SpecRelXXI}) with $H_\Lambda=\sqrt{\Lambda/3}$.\\

DeSitter spacetime is a maximally-symmetric deformation of Minkowski spacetime, in which the familiar Lorentz group in four dimensions $L(c) = SO(1,3)$ will be associated with the de Sitter group $dS(c,H_\Lambda)$, which is the Lorentz group $SO(1,4)$ in five dimensions. Therefore a maximally-symmetric four-dimensional spacetime manifold can be constructed by identifying it both metrically and geometrically with the quotient space $dS(c,H_\Lambda)/L(c)$. This procedure is equivalent to the identification of this spacetime manifold with the de Sitter one $dS(1,3)$ which is the one sheeted four-dimensional hyperboloid of equation
\begin{equation}
\eta_{AB}X^A X^B = (X^0)^2-(X^1)^2-(X^2)^2-(X^3)^2-(X^4)^2=-\frac{1}{H_\Lambda^2}\,,\label{5DdS}
\end{equation}
where $\eta_{AB}$ is the five-dimensional Minkowskian metric with signature $(+,-,-,-,-)$.
Due to the presence of curvature, in general, no privileged class of reference frames exists on the Minkowskian hyperboloid and any choice of a local coordinate system may be in principle acceptable for the description of physical phenomena; however in order to recognize at first sight a solution with exponential scale factor \eqref{scfactds}, we will focus on the flat set of coordinates:
\begin{eqnarray}
X^0&=&\frac{1}{H_\Lambda}\sinh(H_\Lambda x^0)+\frac{H_\Lambda}{2}e^{H_\Lambda x^0}|x|^2\nonumber\\
X^i&=& x^i e^{H_\Lambda x^0}\label{dSflatcoordinates}\\
X^4&=&\frac{1}{H_\Lambda}\cosh(H_\Lambda x^0)-\frac{H_\Lambda}{2}e^{H_\Lambda x^0} |x|^2\,,\nonumber
\end{eqnarray}
Since the line-element should be invariant under any set of coordinates, one can find the relation between the five-dimensional $\eta$ and the four-dimensional $g$ metric imposing
\begin{equation}
\Delta s^2=\int_0^1\eta_{AB}\dot{X}^A(s)\dot{X}^B(s)\, ds=\int_0^1 g_{\alpha\beta}(x)\dot{x}^\alpha(s)\dot{x}^\beta(s)\, ds\, ,\label{dSlinel}
\end{equation}
and therefore
\begin{equation}
g_{\mu\nu}(x)=\eta_{AB}\frac{\partial X^A}{\partial x^\mu}\frac{\partial X^B}{\partial x^\nu}
=\left(\begin{array}{cccc}
1 & 0 & 0 & 0\\
0 & -e^{2 H_\Lambda x^0} & 0 & 0\\
0 & 0 & -e^{2 H_\Lambda x^0} & 0\\
0 & 0 & 0 & -e^{2 H_\Lambda x^0}\\
\end{array}\right)\,.\label{dSmetric}
\end{equation}
In order to find the spacetime symmetry generators $\Pi_0$, $\Pi_i$, ${\cal N}_i$ and ${\cal R}_i$ in one can use the Killing differential equation
\begin{equation}
\partial_\alpha g_{\mu\nu}\Xi^\alpha + g_{\alpha\nu}\partial_\mu\Xi^\alpha + g_{\mu\alpha}\partial_\nu\Xi^\alpha = 0\,.
\end{equation}
Such an equation, using the metric \eqref{dSmetric} reduces to the differential equations system:
\begin{equation}
\left\{\begin{array}{l}
\partial_0\Xi^0=0\\
\partial_i\Xi^0=e^{2 H_\Lambda x^0}\partial_0\Xi^i\\
\partial_i\Xi^i=-H_\Lambda\Xi^0\\
\partial_i\Xi^j +\partial_j\Xi^i=0
\end{array}\right.\;\;\;,\;\;\;
\end{equation}
whose solutions are:
\begin{eqnarray}
\Xi^0(x)&=& a^0 +\delta_{jk}\xi^j x^k\,,\\
\Xi^i(x)&=& -a^0 H_\Lambda x^i + a^i-\epsilon_{ijk}\alpha^j x^k+\\
&+&\xi^i\left(\frac{1-e^{-2 H_\Lambda x^0}}{2H_\Lambda}+\frac{H_\Lambda}{2}|x|^2\right)-H_\Lambda x^i\delta_{jk}\xi^j x^k\,,\nonumber
\end{eqnarray}
where $a^\mu,\alpha^i,\xi^i$ are the spacetime transformations parameters.Once found the Killing vectors explicit expression, using the total charge definition $Q\equiv \Xi^\mu(x) p_\mu$, where the $p_\mu$ are the locally-flat quadri-momenta, it is possible to identify the symmetry generators representations
\begin{eqnarray}
&\Pi_0=p_0-H_\Lambda x^i p_i\;,\;\;\;\Pi_i=p_i\,,\mathcal{R}_i=\epsilon_{ijk}x^j p_k&\label{reldeSPip}\\
&\mathcal{N}_i=x^i p_0+\left(\frac{1-e^{-2 H_\Lambda x^0}}{2H_\Lambda}+\frac{H_\Lambda}{2}|x|^2\right)p_i-H_\Lambda x^i x^k p_k\,,&
\end{eqnarray}
and finally the deSitter group $SO(1, 4)$ with Lie algebra $U(so(1, 4))$:
\begin{eqnarray}
&[\Pi_0,\Pi_i]=-i H_\Lambda \Pi_i\;,\;\;\;[\Pi_i,\Pi_j]=0\;,&\label{momentadScomm}\\
&[\mathcal{R}_i,\Pi_0]=0\;,\;\;\; [\mathcal{R}_i,\Pi_j]=i\epsilon_{ijk}\Pi_k\;,\;\;\;[\mathcal{R}_i,\mathcal{R}_j]=i\epsilon_{ijk}\mathcal{R}_k&\\
&[\mathcal{N}_i,\Pi_0]=-i\left(\Pi_i-H_\Lambda{\cal N}_i\right)\;,\;\;\;[\mathcal{N}_i,\Pi_j]=-i\Pi_0\delta_{ij}-i H_\Lambda \epsilon_{ijk}\mathcal{R}_k\;,&\\
&[\mathcal{R}_i,\mathcal{N}_j]=i\epsilon_{ijk}\mathcal{N}_k\;,\;\;\;[\mathcal{N}_i,\mathcal{N}_j]=-i\epsilon_{ijk}\mathcal{R}_k\;.&
\end{eqnarray}
Whose first Casimir is
\begin{equation}
{\cal H}=\left(\Pi_0+x^i\Pi_i\right)^2-e^{-2 H_\Lambda x^0}|\Pi|^2=p_0^2-e^{-2 H_\Lambda x^0}|p|^2\,.\label{dSIcas}
\end{equation}
It is also possible at this point to characterize spacetime coordinates transformations under boost, through the differential equations arising from the relation
\begin{equation}
\frac{d x^\mu}{d\xi^i}=-i[\mathcal{N}_i,x^\mu]\,.
\end{equation}
With the right choice of coordinates, imposing to be aligned with the point to be boosted, {\it id est} imposing for instance $i=1$ and the point original coordinates to be $(\bar{x}^0,\bar{x}^1,0,0)$, after some cumbersome calculations, one can find:
\begin{equation}
\left\{\begin{array}{l}
x^0(\xi)= \frac{1}{H_\Lambda}\ln\big(e^{H_\Lambda\bar{x}^0}\big((\cosh\frac{\xi}{2} - H_\Lambda\bar{x}^1\sinh\frac{\xi}{2})^2 -e^{-2 H_\Lambda \bar{x}^0} \sinh^2\frac{\xi}{2}\big)\big)\\
x^1(\xi)=\frac{e^{2\xi}-1-e^{2H_\Lambda\bar{x}^0}e^{-\xi}(1-H_\Lambda\bar{x}^1)^2+e^{2H_\Lambda\bar{x}^0}(1+H_\Lambda\bar{x}^1)^2}{H_\Lambda  \left(e^{2H_\Lambda\bar{x}^0} (1+H_\Lambda\bar{x}^1+e^{\xi} (1-H_\Lambda\bar{x}^1)^2)-(e^\xi-1)^2\right)}
\end{array}
\right.\,.\label{dSspacetimetrasf}
\end{equation}
The same procedure for quadri-momenta gives
\begin{equation}
\left\{\begin{array}{l}
p_0(\xi)=-\bar{p}_0\cosh\left( \ln\left(\frac{1-e^\xi- e^{H_\Lambda\bar{x}^0}\left(1+H_\Lambda\bar{x}^1+e^\xi(1-H_\Lambda\bar{x}^1)\right)}{1-e^\xi+ e^{H_\Lambda\bar{x}^0}\left(1+H_\Lambda\bar{x}^1+e^\xi(1-H_\Lambda\bar{x}^1)\right)}\right) \right)+\\
\hspace{1cm}+e^{-H_\Lambda\bar{x}^0}\bar{p}_1\sinh\left( \ln\left(\frac{1-e^\xi- e^{H_\Lambda\bar{x}^0}\left(1+H_\Lambda\bar{x}^1+e^\xi(1-H_\Lambda\bar{x}^1)\right)}{1-e^\xi+e^{H_\Lambda\bar{x}^0}\left(1+H_\Lambda\bar{x}^1+e^\xi(1-H_\Lambda\bar{x}^1)\right)}\right) \right)\\
p_1(\xi)=\frac{e^{-\xi}e^{-2H_\Lambda\bar{x}^0}}{4}\big((e^\xi-1)^2\bar{p}_1-e^{2H_\Lambda\bar{x}^0}\big(1+H_\Lambda\bar{x}^1+ e^\xi(1-H_\Lambda\bar{x}^1)\big)\times\\ \hspace{1cm}\times\big(2(e^\xi-1)\bar{p}_0-\bar{p}_1\big(1+H_\Lambda\bar{x}^1+e^\xi(1-H_\Lambda\bar{x}^1)\big)\big)\big)
\end{array}
\right.\,.\label{dSmomentatrasf}
\end{equation}
Both \eqref{dSspacetimetrasf} and \eqref{dSmomentatrasf} in the limit $H_\Lambda\rightarrow 0$ reduce to the classical Lorentz transformations
\begin{eqnarray}
&&\left\{\begin{array}{l}
x^0(\xi)=\bar{x}^0\cosh(\xi)-\bar{x}^1\sinh(\xi)\\
x^1(\xi)=-\bar{x}^0\sinh(\xi)+\bar{x}^1\cosh(\xi)
\end{array}
\right.\,,\\
&&\left\{\begin{array}{l}
p_0(\xi)=\bar{p}_0\cosh(\xi)-\bar{p}_1\sinh(\xi)\\
p_1(\xi)=-\bar{p}_0\sinh(\xi)+\bar{p}_1\cosh(\xi)
\end{array}
\right.\,.
\end{eqnarray}
It is pretty easy to verify that both the line element obtained from \eqref{dSlinel}:
\begin{equation}
\Delta s^2 = \frac{4}{H_\Lambda^2} \sinh\left(\frac{H_\Lambda}{2} x^0\right)^2 - e^{H_\Lambda x^0} |x|^2\,,
\end{equation}
and the first Casimir ${\cal H}$ introduced in \eqref{dSIcas}, are invariant under the $H_\Lambda$-deformed Lorentz transformations \eqref{dSspacetimetrasf} and \eqref{dSmomentatrasf}.\\
Long story short, deSitter group is a General Relativity induced deformation to Poincar\'{e} group. Other kinds of Poincar\'{e} algebra deformations have been applied to the study of spacetime quantization. Such models are known as Doubly Special Relativity (DSR) models \cite{AmelinoDSR2001,AmelinoDSR2002,SmMagDSR2001,SmMagDSR2003,AmelinoDSR2010}, they are mostly based on the so called $\kappa$-Poincar\'{e} Hopf algebra.

\section{Hopf Algebras for dummies}
\noindent 

The most significant difference between Hopf algebras and the more familiar Lie algebras, is the presence of a non-trivial coproduct, which is an element of the coalgebric sector \cite{Fuchs,MajidQGP}. While an algebra is a vector space on a field $F$, in which we have defined the concept of:
\begin{equation}
\begin{array}{ll}
Product\;\;\; & {\cal M}:{\cal A}\otimes{\cal A}\rightarrow {\cal A}\\
Unity\;\;\; & \eta:F\rightarrow {\cal A}
\end{array}
\end{equation}
a coalgebra is a vector space ${\cal A}$ on a field $F$, provided of the linear maps:
\begin{equation}
\begin{array}{ll}
Coproduct\;\;\; & \Delta:{\cal A}\rightarrow {\cal A}\otimes{\cal A}\\
Counity\;\;\; & \epsilon:{\cal A}\rightarrow F
\end{array}
\end{equation}
The coalgebras theory is essentially dual to the one of the algebras. In fact, if ${\cal A}$ is a coalgebra, then is dual space ${\cal A}*$ (the space of the linear functionals $\Phi : {\cal A}\rightarrow F$) inherits a punctual algebra structure induced by $\Delta$ and $\epsilon$.\\
An Hopf algebra is a vector space ${\cal A}$  on a field $F$ which is simultaneously both an algebra and a coalgebra, and which is provided by a further linear map called {\it antipode}:
\begin{equation}
{\cal S}:{\cal A}\rightarrow {\cal A}
\end{equation}
whose role is analogous to the inverse element in the link between an algebra and a coalgebra. A Lie algebra can be generalized into an Hopf algebra by mean of its {\it universal enveloping algebra} $U(g)$, trivially defining coproduct, counity and antipode:
\begin{equation}
\begin{array}{l}
\Delta (X) = X\otimes \mathbb{1} + \mathbb{1}\otimes X \\
\epsilon(X)=0\\
{\cal S}(X)=-X
\end{array}\label{trivialcopr}
\end{equation}
Hopf algebras are a very powerful tool for the study of Planck-scale physics, in fact, interpreting the Planck length $L_P=\sqrt{\hbar G/c^3}$ as an observer-independent scale rather than a coupling constant, might bring to a direct conflict with the basic principles of Special Relativity, in which energies and lengths cannot be observer-independent. To overcome this issue one can rely to a theory with two observer-independent scales, the already mentioned DSR models.\\

\subsection{$\kappa$-Poincar\'{e}}

The Deformed Special Relativity approach to Quantum Gravity assumes the possibility to formalize the spacetime quantization through some sort of coordinates noncommutativity, such as
\begin{equation}
[\chi^\mu,\chi^\nu]=\Theta^{\mu\nu}+\Xi_\gamma^{\mu\nu}\chi^\gamma\,.
\end{equation} 
The $\Theta^{\mu\nu}\neq 0$ case is the so called "canonical" approach to coordinates noncommutativity, while specializing on the case $\Xi_\gamma^{\mu\nu}\sim\ell$ one obtains the $\kappa$-Minkowski coordinates commutation rules:
\begin{equation}
[\chi^0,\chi^i]=-i\ell \chi^i\label{kMink}\,,
\end{equation}
where $\ell$ is a parameter with the dimensions of a length that we expect to be of the order of the inverse of Planck scale $\ell\sim 1/M_P$.\\
The request for the relation \eqref{kMink} to be invariant under a 10-generators spacetime symmetry group, imposes to take into account an $\ell$-deformed algebra, as extensively discussed in \cite{MajidRuegg}. Given the parallelism between \eqref{kMink} and the first relation of \eqref{momentadScomm}, the most studied deformed algebras taken into account are deSitter and ant-deSitter inspired Hopf algebras (see for instance \cite{LukNowRueggTol,Lukierski:1992dt,Luk1993,Luk1994,KowaNowakdS})).\\

The natural candidate for this theory's algebra of symmetries is a family of quantum (Hopf) algebras generalizing the standard Poincar\'{e} algebra. In particular the DSR proposal can be realized in the framework of quantum $\kappa$-Poincar\'{e} algebra \cite{LukDSR,KowalskiDSR}. This algebra contains a natural deformation parameter of mass $\kappa\sim M_P$ by construction (for our purposes we will prefer to consider as deformation parameter a length scale $\ell$, defined as $\ell= 1/\kappa$).\\
\\
$\kappa$-Poincar\'{e} Hopf algebra {\it bicrossproduct basis} is characterized \cite{MajidRuegg} by the following commutation relations:
\begin{eqnarray}
&[p_i,p_j]=0\,,\;\;\;[{\cal R}_{i},{\cal R}_{j}]=-i\epsilon_{ijk}{\cal R}_{k}\;,\;\;\;[{\cal R}_{i},p_j]=-i \epsilon_{ijk} p_k\,,\nonumber&\\
&\left[{\cal N}_{i},p_0\right]=-i p_i\;,\;\;\;
[{\cal N}_{i},p_j]=-i \delta_{ij}\left(\frac{1-e^{-2\ell p_0}}{2\ell}+\frac{\ell}{2}|p|^2\right)+i\ell p_i p_j\,,&\\
&[{\cal N}_{i},{\cal N}_{j}]=i\epsilon_{ijk}{\cal R}_{k}\,,\;\;\;[{\cal R}_{i},{\cal N}_{j}]=-i\epsilon_{ijk}{\cal N}_{k}\,,&\nonumber
\end{eqnarray}
and the following coproducts and antipodes
\begin{eqnarray}
&\Delta p_0 = p_0\otimes \mathbb{1} + \mathbb{1}\otimes p_0\,,&\label{encopr} \\
&\Delta p_i = p_i\otimes \mathbb{1} + e^{-\ell p_0}\otimes p_i\,,&\label{momcopr}\\
&{\cal S}(p_0)=-p_0\,,\;\;\;{\cal S}(p_i)=-e^{\ell p_0}p_i\,,&\\
&\Delta {\cal N}_{i} ={\cal N}_{i}\otimes \mathbb{1} + e^{-\ell p_0}\otimes {\cal N}_{i}-i\ell\epsilon_{ijk} p_j\otimes {\cal R}_{k}\,,&\label{boostcopr}\\
&{\cal S}({\cal N}_{i})=-e^{\ell p_0}{\cal N}_{i}+i\ell\epsilon_{ijk} p_j {\cal R}_{k}\,,&
\end{eqnarray}
where the ${\cal R}_{i}$ are the undeformed rotations and ${\cal N}_{i}$ is the boost generator which can be represented as
\begin{eqnarray}
{\cal N}_{i}&=&\chi^0 p_i + \chi^i \left(\frac{1-e^{-2  \ell p_0}}{2 \ell }+\frac{\ell}{2}  
|p|^2\right)=x^0 p_i + x^i \left(\frac{1-e^{-2  \ell p_0}}{2 \ell }+\frac{\ell}{2}  
|p|^2\right)- \ell x^j p_j p_i\,.   \nonumber
\end{eqnarray}
The boost can be both represented in terms of noncommuattive $\chi^\alpha$ or commutative coordinates $x^\alpha$, the map linking the two coordinatizations 
\begin{equation}
\chi^0=x^0-\ell x^i p_i\,,\;\;\;\chi^j=x^j\,,\label{commnoncommmap}
\end{equation}
is well known in the literature (see {\it exempli gratia} \cite{kBob,SpecRelLoc}) and often used to work with an undeformed symplectic sector $[p_\alpha,x^\beta]=-i\delta_\alpha^\beta$. The duality between noncommutative coordinates and the deSitter physical momenta $\Pi_\mu$ has been explored in \cite{Lateshift}, finding that the map \eqref{commnoncommmap} has the same form as the one relating local momenta $p_\mu$ with the conserved charges \eqref{reldeSPip} that one can find solving the deSitter Killing equation.

The interpretation of the physical consequences of this kind of deformed algebra is still debated, in some cases they seem to imply features like particles in-vacuo dispersion ({\it id est} momentum-dependent photon velocity) \cite{AmelinoDSR2001,AmelinoDSR2002,AmelinoDSR2010,Bob,kBob} and deformed non-linear composition laws for momenta, strictly connected with the coproduct nontriviality.\\
The relative locality interpretation \cite{RelLoc1,RelLoc2} portrays those features as momentum-space curvature effects: the momentum-dependent massless particles velocity has been succesfully described \cite{Lateshift} as some sort of dual redshift effect or {\it lateshift}, while the modified composition law is in general identified at first orders as 
\begin{equation}
(p\oplus q)_\alpha\equiv p_\alpha +q_\alpha - \ell Z_\alpha^{\beta\gamma}p_\beta q_\gamma +\dots\label{Freidcomp}
\end{equation}
where the physical interpretation of $Z$ have not been univocally identified and is still debated\footnote{It have been often postulated the identification between $Z$ and the momentum-space connections $Z_{\alpha}^{\beta\gamma}\equiv\Gamma_{\alpha}^{\beta\gamma}$, however such an assumption imposes to take into account some kind of momentum-space torsion \cite{RelLoc1,GiuFla}, which we won't take into account here, since deSitter spacetime has no torsion.}.\\
 Moreover another feature characterizes the modified composition law, namely the {\it backreaction} \cite{Majidback,GiuFla,CarCorFla,FlaNiccoMel}. This is a property of the composition law under boost: if we have a particle with four-momentum $k_\alpha$ decaying into two particles with momenta $p_\alpha$ and $q_\alpha$, one would like that the composition law observed in a boosted reference frame with boost parameter $\xi$ to have the same form as the one at rest, however:
\begin{equation}
k'_\alpha(\xi)\neq (p'(\xi)\oplus q'(\xi))_\alpha\,.
\end{equation} 
The reason for this weird behavior is that in those models momenta "backreact" on the boost parameter itself $\xi\vartriangleleft (p,q)$ \cite{GiuPalm}, according to the boost nontrivial coproduct \eqref{boostcopr}.
This last aspect of $\kappa$-Poincar\'{e} algebra may be a little tricky to explain, however, since the {\it lateshift} approach is so effective in describing Hopf algebra features just as momentum-space curvature effects with a deSitter-like metric, with no need for torsion or nonmetricity, why not to try a similar strategy to formalize the modified composition law and the backreaction?\\
As long as one can find a duality relation between time delay effects and redshift, it may be possible also to find a similar liaison between  interacting particles in an $\ell$ deformed momentum-space and observers in deSitter spacetime.
The identification of a satisfying physical interpretation of such features using deSitter spacetime as inspiration is the heart of this article and will be extensively addressed in sec. \ref{sec:Composition1}. Before elaborating any further about the Hopf-algebraic aspects of deSitter geometry we may need to clarify a little bit more how $\kappa$-Poincar\'{e} momentum-space and deSitter spacetime are related by reviewing some well established DSR features and their physical interpretation.

 \subsection{Parallelism between $\kappa$-Poincar\'{e} phenomenology and deSitter Redshift}\label{sec:kPphysics}
 \noindent
 
 $\kappa$-Poincar\'{e} curved momentum-space algebra belongs to a well populated family of quantum groups of spacetime symmetries \cite{FlaLizziManfredonia}. In this case we will focus on the deSitter-like geometry, starting from \cite{Trevisan,GiuPalm,Barcaroli:2016yrl} the Hamiltonian obtained by integrating the momentum-space invariant line-element
 	\begin{equation}
 	{\cal H}=\int_0^1 ds\,\zeta^{\alpha\beta}(P)\dot{P}_\alpha\dot{P}_\beta\,,\label{intCasimir}
 	\end{equation}
 	where the $\zeta$ is the momentum-space metric and where the momentum-space geodesics $P(s)$ are defined by the geodesic equation: 
 	\begin{equation}
 	\ddot{P}_\alpha + \Gamma_\alpha^{\beta\gamma}\dot{P}_\beta\dot{P}_\gamma= 0\,.\label{geochristoffel}
 	\end{equation}
 	The deformed Hamiltonian \eqref{intCasimir} is related to the deSitter-like momentum-space metric
 	\begin{equation}
 	\zeta^{\alpha\beta}=\left(\begin{array}{cccc}
 	1&0&0&0\\
 	0&-e^{-2\ell P_0}&0&0\\
 	0&0&-e^{-2\ell P_0}&0\\
 	0&0&0&-e^{-2\ell P_0}
 	\end{array}\right)\,,
 	\end{equation}
 	through the on-shell relation ${\cal H}=m^2$ where the mass is the invariant momentum-space line-element defined as
 	\begin{equation}
 	m^2=\int_0^1 \zeta^{\alpha\beta}(P(s))\dot{P}_\alpha(s)\dot{P}_\beta(s)\,ds\,,
 	\end{equation} 
 	where the $P(s)$ are momentum-space geodesics which satisfy $P_\alpha(0)=0\,, P_\alpha(1)=p_\alpha$ and whose explicit expressions
 	\begin{eqnarray}
 	P_0(s)&=&-\frac{1}{\ell} \ln (1 + (e^{\ell p_0} - 1) s)\,,\\
 	P_i(s)&=&\frac{e^{\ell p_0} p_i s}{1 + (e^{\ell p_0} - 1) s}\,,
 	\end{eqnarray}
 	can be obtained solving the geodesic equations \eqref{geochristoffel}, in which the connection $\Gamma$ is the usual Christoffel symbol in momentum-space
 	\begin{equation}
 	\Gamma_\lambda^{\mu\nu}=\frac{1}{2}\zeta_{\lambda\sigma}\left(\frac{\partial}{\partial p_\mu}\zeta^{\sigma\nu}+\frac{\partial}{\partial p_\nu}\zeta^{\sigma\mu}-\frac{\partial}{\partial p_\sigma}\zeta^{\mu\nu}\right)\,,
 	\end{equation}
 	with no torsion or nonmetricity:
 	\begin{eqnarray}
 	&\Gamma_0^{00}=\Gamma_0^{01}=\Gamma_0^{10}=\Gamma_i^{00}=\Gamma_i^{ii}=0\,&\label{Christ1}\\
 	&\Gamma_0^{ii}=-\ell e^{-2\ell p_0}\;,\;\;\;\Gamma_1^{10}=\Gamma_i^{0i}=-\ell \,.\label{Christ2}&
 	\end{eqnarray}
 
We can then write down the Hamiltonian as the algebra first Casimir:
\begin{equation}
{\cal H}=\frac{4}{\ell^2}\sinh^2\left(\frac{\ell}{2}p_0\right)-e^{\ell p_0}|p|^2\,,\label{Hamiltonian}
\end{equation}
this leads to a modified Hamiltonian framework, characterized by a rich phenomenology of deformation effects for a free massless particle propagating in spacetime, formalized by the (classical) Hamilton equations:
\begin{equation}
\dot{x}^\beta=\{{\cal H},x^\beta\}\,.
\end{equation}
Using this relation it easy to find the explicit worldlines expression:
\begin{equation}
x^1-\bar{x}^1=\int_{\bar{x}^0}^{x^0}\frac{dx^1}{dx^0}dx^0=\int_{\bar{x}^0}^{x^0}\frac{\dot{x}^1}{\dot{x}^0}dx^0=e^{\ell p_0}(x^0-\bar{x}^0)\,.\label{kPworldli}
\end{equation}
If we imagine to have two massless particle emitted at the same time at an observer's (Alice) origin $(x^0_A=0,x^1_A=0)$ with different energies: an high energetic $p_0^h$ and a low energetic one $p_0^l$, the worldlines \eqref{kPworldli} imply that a translated observer (Bob) will not receive those two particles at the same time. In fact in Bob's frame
\begin{equation}
x^\alpha_B\simeq x^\alpha_A-a^\beta\{p_\beta , x^\alpha\}\, ,\label{coordtrans}
\end{equation}
where $(a^0,a^1)$ are the translation coefficients, the two massless particles will arrive at the spacial origin $x^1=0$ at two different times, whose difference characterizes the time delay
\begin{equation}
\Delta x_B^0=a^0 (e^{\ell (p_0^h- p_0^l)}-1)\simeq \ell a^0(p_0^h-p_0^l)\,.
\end{equation}
 Relative Locality \cite{RelLoc1,RelLoc2,Bob,kBob} is a classical framework in which both quantum and spacetime curvature  effects are turned off, i.e. $\hslash\rightarrow 0$ and $G\rightarrow 0$. At the same time however Planck-scale   deformations are still present as modification to the classical Hamiltonian mechanics since in this limit $\ell\sim M_P=\sqrt{G/\hbar}\neq 0$. In this limit then all the aforementioned Hopf-algebraic features are treated as remnants of a quantum regime in a classical framework with deformed Hamiltonian dynamics with a curved momentum-space. For this reason all the features so far formalized using commutation relations can be also  expressed as nontrivial Poisson brackets, following the well known relation $\left[A,B\right]=i\hslash \{A,B\}$ and the coordinate set taken into account will be the locally flat commutative one $x^\alpha$ instead of the noncommuative one $\chi^\beta$.

 \subsection{Lateshift}
 \noindent

The duality between the $\kappa$-Poincar\'{e} time-delay and the spacetime redshift effect in deSitter framework is pretty easy to point out \cite{Lateshift}. First of all one can notice that given \eqref{Hamiltonian} and \eqref{kPworldli}, in the massless case we have
\begin{equation}
\begin{array}{lccc}
\; & \text{Worldlines} &\;& \text{Hamiltonian}\\
\text{deSitter} & \Delta x^1=\frac{1-e^{-H_\Lambda \Delta x^0}}{H_\Lambda} &\;& |p_1|=e^{H_\Lambda x^0} p_0\\
\;&\;&\;&\;\\
\kappa\text{-Poincar\'{e}} & \Delta x^1=e^{\ell p_0}\Delta x^0 &\;& |p_1|=\frac{1-e^{-\ell p_0}}{\ell}
\end{array}\, .
\end{equation}
The Relativity of locality of $\kappa$-Minkowski noncommutative spacetime model can be defined observing that in such a framework events that are coincident and occur in the origin of a given observer Alice are objectively coincident for all observers that share Alice's origin (all observers which are
purely boosted with respect to Alice), but may not appear to be coincident to an observer Bob far from Alice. Locality is still objective, but the abstraction of "distant locality" is lost.\\
According to \eqref{kPworldli}, two photons with energy $\bar{p_0}$ and $\tilde{p}_0$ emitted at Alice origin $(\bar{x}^0_A=0,\bar{x}^1_A=0)$ will propagate along the worldlines 
\begin{equation}
x^1_A=e^{\ell \bar{p}_0} x^0_A\;,\;\;|;x^1_A=e^{\ell \tilde{p}_0} x^0_A\,.
\end{equation}
Given the coordinate transformations \eqref{coordtrans} between Alice and Bob
\begin{equation}
\bar{x}^0_B=\bar{x}^0_A-a^0\;,\;\;\;\bar{x}^1_B=\bar{x}^1_A-a^0\,,
\end{equation}
and the relation between the translation parameters $a^1=e^{\ell \bar{p}_0}a^0$ (which can be found defining Bob's origin as the intersection between Bob and the first worldline), Bob will express the two worldlines as
\begin{eqnarray}
x^1_B=e^{\ell \bar{p}_0} (x^0_B-a^0)-e^{\ell \bar{p}_0}a^0\;,\;\;\;x^1_B=e^{\ell \tilde{p}_0} (x^0_B-a^0)-e^{\ell \bar{p}_0}a^0\,.
\end{eqnarray}
Therefore the time delay between the arrival af the two particles at Bob's spatial origin ($x^1_B=0$, where the detector lies) can be expressed as
\begin{equation}
\frac{\Delta x^0_B}{\bar{x}^0_B}=1-e^{-\ell \Delta p_0}\,,
\end{equation}
where $\Delta p_0=\tilde{p}_0-\bar{p}_0$.\\
The duality between this effect and the redshift in momentum-space can be highlighted by asking what happens in this framework in an exact specular situation: if in $\kappa$-Poincar\'{e} we are interested in the time delay arising between two particles with different energies emitted at the same time, in the deSitter case we will take into account two photons emitted at different times with the same energy. 
Let's consider then a first photon emitted with a generic energy $p_0^A$ from some standard candle (Alice), being observed by a translated observer Bob at a distance $(a^0,a^1)$. The energy of the photon in Bob's frame will be determined by the well known redshift formula in deSitter:
\begin{equation}
\bar{p}_0^B=e^{- H_\Lambda a^0}p_0^A\,.
\end{equation}
 After some time $\Delta x^0$, in an expanding spacetime the distance between the two observers will of course increase  and the energy $\tilde{p}_0^B$ measured by Bob will be further redshifter by
\begin{equation}
\tilde{p}_0^B=e^{-H_\Lambda (a^0+\Delta x^0)}p_0^A=e^{- H_\Lambda \Delta x^0}\bar{p}_0^B\,.
\end{equation}
Therefore the redshift effect can be characterized as
\begin{equation}
\frac{\Delta p_0^B}{\bar{p}_0^B}=1-e^{- H_\Lambda \Delta x^0}\,.
\end{equation}
{\it Quod erat demonstrandum}. It is worth to mention that in order to find this duality between the $\kappa$-Poincar\'{e} momentum-space and deSitter spacetime we didn't have to assume any torsion or nonmetricity in the geometry of our manifolds, on the other hand looking for an expression for the deformed momentum composition law in terms of parallel-transported vectors may lead to more complex scenarios that we may not need in a lateshift-like approach.

\section{Nontrivial composition law for momenta and deSitter reference frames}\label{sec:Composition1}

While the deformed Hamiltonian \eqref{Hamiltonian} formalizes the $\kappa$-Poincar\'{e} modified dynamics of particles, the coproducts such as \eqref{momcopr} and \eqref{encopr} give us some indications about the  particles' interaction. In fact the coproduct can also be expressed \cite{MelSum,FlaNiccoMel} as 
\begin{equation}
\Delta p_\mu={\cal D}_\mu(p\otimes \mathbb{1},\mathbb{1}\otimes p)\,,\label{coprcomp}
\end{equation} 
where the function ${\cal D}_\mu(p,q)$ describes a deformed composition of momenta ${\cal D}_\mu(p,q)=(p\oplus q)_\mu$.\\
In other words, since translations in $\kappa$-Minkowski spacetime, defined from the $\kappa$-Minkowski commutation rules between coordinates \eqref{kMink} can be represented \cite{Lukierski:1992dt,LukNowRueggTol} through simple derivatives:
\begin{equation}
p_0\equiv -i\frac{\partial}{\partial \chi^0}\;,\;\;\; p_0\equiv -i\frac{\partial}{\partial \chi^0}\,,
\end{equation}
we can obtain the total momentum and energy of two interacting particles, calculating the the eigenvalues of those operators acting on two (time-to-the-right) plane waves $\Phi(\chi)=e^{i p_i\chi^i}e^{i p_0\chi^0}$ and $\Psi(\chi)=e^{i q_i\chi^i}e^{i q_0\chi^0}$ .\\
Since our coordinates are noncommutative we should make use of the Baker-Campbell-Hausdorff formula which implies that in case $[X,Y]=s Y$ then $e^{X}e^Y=e^{e^s Y}e^X$. Moving all the time coordinates to the right we have that one can define the composition rule for the particles' momenta as
\begin{equation}
e^{i p_i \chi^i}e^{i p_0 \chi^0}e^{i k_i \chi^i}e^{i k_0 \chi^0}=e^{i (p\oplus k)_i \chi^i}e^{i (p\oplus k)_0 \chi^0}\,,
\end{equation} 
as discussed in \cite{AmelinoMajid}. Therefore since we have that
\begin{equation}
[p_0 \chi^0,k_i \chi^i]=[p_0,k_i\chi^i]\chi^0+p_0[\chi^0,k_i\chi^i]=p_0 k_i[\chi^0,\chi^i]=p_0 k_i(-i\ell \chi^i)\,,
\end{equation}
using the BKH-formula we can identify the composition rules of this model as
\begin{equation}
k_\mu=\left\{\begin{array}{l}
(p\oplus q)_0=p_0+q_0\\
(p\oplus q)_i=p_i+e^{-\ell p_0}q_i
\end{array}\right.\,,\label{vertice}
\end{equation}
which is in agreement with the definition \eqref{coprcomp}.\\
This kind of energy and momenta composition rules modifications have in general striking implications for the macroscopical limit which are currently not entirely comprehended. The discussion on how to compose those effect through some sort of Planck-scale-sensitive statistical mechanics have lead to the so called {\it soccerball problem} \cite{HinterSoccerball,AmelinoSoccerball} which tries to address the issue of macroscopic manifestations which of course are not detected in table-top experiments. There is some literature about how to solve the soccerball problem using curved momentum-space formalism \cite{RelLocSoccerball}. In this paper we would like to give a few more insight about such an interpretation, showing how an explicit duality can be identified between modified composition laws for momenta in $\kappa$-Poincar\'{e} and deSitter spacetime features.

\subsection{Composition law in deSitter spacetime}\label{sec:OplusdSspacetime}

\begin{figure}[h!]
 	\centering
\framebox{\includegraphics[scale=0.35]{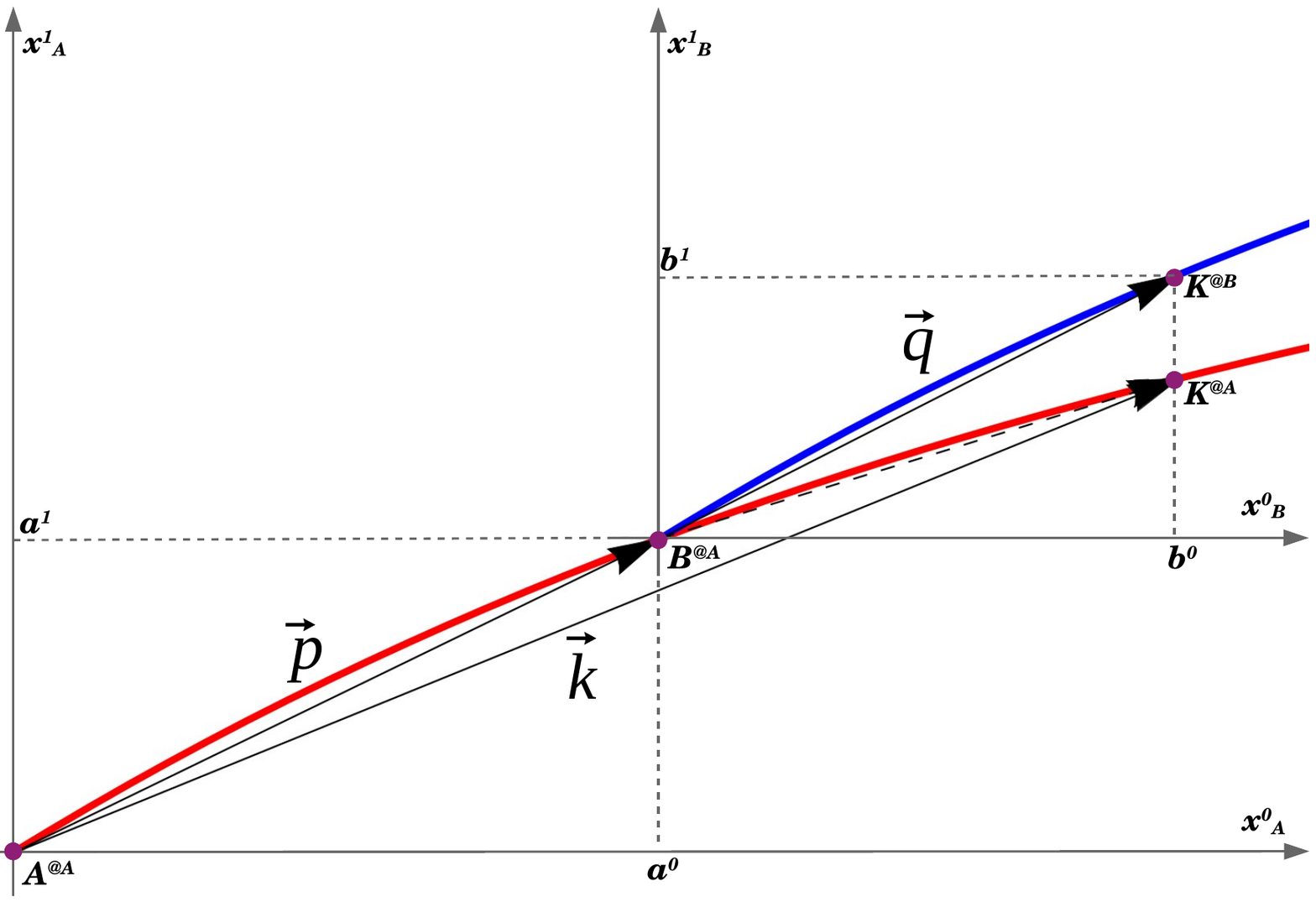}}
\caption{\footnotesize Alice tries to express the components of vector $\vec{k_{@A}}$ in terms of vector $\vec{p}_{@A}$ which lies in her reference frame and $\vec{q}_{@B}$ whose components are defined by the translated observer Bob.}
\label{fig:ABdeSitter}
\end{figure}

As well as it is possible to define a strict duality between redshift and the time-delay effect, it is also possible to explain $\kappa$-Poincar\'{e} non trivial composition law of momenta as a composition of vectors in deSitter spacetime, however since the boost acts on the composition law nonlocally, {\it id est} one needs to take into account the backreaction effect to ensure the invariance of interaction vertices under boost transformations, we won't assume this composition law to be the sum of parallel transported vectors as in subsection \ref{Freidcomp}. We can imagine the two vectors to lie at the origins of two translated observers Alice and Bob, which are connected by a photon worldline
\begin{equation}
x_A^1(x_A^0)=\frac{1-e^{-H_\Lambda x^0_A}}{H_\Lambda}\label{dSworldline}
\end{equation}
 so that both their origins sit in different point of such worldline. Let's now say that Alice wants to know what are the coordinates of some vector $\vec{k}$ that in terms of Bob coordinates is $\vec{q}_{@B}=(b^0,b^1)$, knowing that their origins are connected by some other vector that Alice expresses as $\vec{p}_{@A}=(a^0,a^1)$ (see Figure \ref{fig:ABdeSitter}). 
How can Alice formalize the vector $\vec{k}_{@A}$ as the composition of the coordinates of $\vec{p}_{@A}$ and $\vec{q}_{@B}$?
For the class of observers reached in their spacetime origin by the signal emitted in Alice's spacetime origin $(x^1_B(x^0_B=0)=0)$, it is possible to define the following coordinate transformations under translation:
\begin{eqnarray}
x^0_B&=&\bar{x}^\alpha\Pi_\alpha\vartriangleright x^0_A=x^0_A -a^0\\
x^1_B&=&\bar{x}^\alpha\Pi_\alpha\vartriangleright x^1_A=e^{H_\Lambda a^0}\left(x^1_A-\frac{1-e^{-H_\Lambda a^0}}{H_\Lambda}\right)\,,
\end{eqnarray}
where the action of the symmetry generators $\Pi_0=p_1 -H_\Lambda x^1 p_1$, $\Pi_1=p_1$ on the phase space functions is represented through the ordinary action-to-the-right of Lie groups
\begin{equation}
\bar{a}{\mathcal G}\vartriangleright A=\sum_{n=0}^{\infty}\frac{1}{n!}\{\bar{a}{\mathcal G},\{\ldots\{\bar{a}{\mathcal G},A\}\ldots\}\,,\label{GenAct}
\end{equation}
in which the $\bar{a}$ are the coordinate transformation parameters.\\
It is then easy to find how Alice will have to transform $\vec{q}_{@B}$ coordinates provided by Bob:
\begin{eqnarray}
k^0_{@A}&=& a^0+b^0\\
k^1_{@A}&=&e^{-H_\Lambda a^0}b^1+\frac{1-e^{-H_\Lambda a^0}}{H_\Lambda}\,.
\end{eqnarray}
Moreover, since Bob's origin sits on the worldline that originates in Alice's one, we can use (\ref{dSworldline}) to find how Alice will express vector $\vec{k}_{@A}$ in terms of $\vec{p}_{@A}$ and $\vec{q}_{@B}$ components:
\begin{eqnarray}
\vec{k}_{@A}&=&\vec{p}_{@A}\oplus\vec{q}_{@B}=(a^0,a^1)\oplus (b^0,b^1)=\vec{p}_{@A}+\vec{q}_{@A}=\nonumber\\
&=&\left(a^0+b^0\,,\,a^1+b^1 e^{-H_\Lambda a^0}\right)\,.\label{kcomposed}
\end{eqnarray}
We can observe that (\ref{kcomposed}) reproduces in spacetime the composition rules \eqref{vertice} expressed for $\kappa$-Poincar\'{e} momentum space.
We don't have in this case an explicit expression in terms of connections as in \cite{GiuPalm,RelLoc1} but a much simpler formalism that only depends on the metric of spacetime. We don't need to explicitly parallel transport any vector but we express $\vec{k}_{@A}$ directly as the sum of vectors $\vec{p}_{@A}$ in Alice's frame  and $\vec{q}_{@B}$ in Bob's one.\\
 Therefore while on the one hand there's a well established duality between worldlines in deSitter and Casimir in $\kappa$-Poincar\'{e}, as well as between redshift and DSR time-delay effect, we have now found that, on the other hand, different particles in $\kappa$-Poincar\'{e} behave like different reference frames in deSitter spacetime.\\

\section{Backreaction in $\kappa$ Poincar\'{e}}

Let's now on $\kappa$-Poincar\'{e} peculiar relativistic features. In relativistic particle physics it is, of course, of paramount importance to coherently formalize how different observers will describe some particles interaction and which are the relations between the observables that they will respectively measure. For instance in the simple case described in \eqref{vertice}, one would expect that since $k_\alpha=(p\oplus q)_\alpha$, then a boosted observer would have $k_\alpha '=(p'\oplus q')_\alpha$, where we have denoted with $k',p',q'$ the momenta measured by the latter. However in the $\kappa$-Poincar\'{e} framework this relativistic prescription is not as intuitive, since:
\begin{equation}
k'_\alpha(\xi)\neq (p'(\xi)\oplus q'(\xi))_\alpha\,\label{neqback}.
\end{equation}
In fact, letting the boost in the direction 1 act as
\begin{eqnarray}
A'&=&\mathcal{N}_{(1)}\vartriangleright A=A+\xi_1\{\mathcal{N}_{(1)},A\}+\frac{\xi_1^2}{2!}\{\mathcal{N}_{(1)}, \{\mathcal{N}_{(1)},A \} \}+\nonumber\\
&+&\frac{\xi_1^3}{3!}\{\mathcal{N}_{(1)},\{\mathcal{N}_{(1)},\{\mathcal{N}_{(1)},A\}\}\}+...\,,
\end{eqnarray}
it is not difficult to verify that the $\kappa$ Poincar\'{e} boost acts differently on the right and on the left term of relation \eqref{neqback}. Since in the following section we will be interested mostly by the non-ultrarelativistic regime of the deSitter transformations \eqref{dSspacetimetrasf}, for sake of simplicity we can here show this different behavior at first order in the rapidity parameter $\xi_1$. Therefore, while the boost of the momenta composition is
\begin{eqnarray}
k'_1&=&k_1 + \xi_1\left(\frac{1 - e^{-2 \ell k_0 }}{2 \ell} - \ell k_1^2  + \frac{\ell}{2} |k|^2 \right)=\nonumber\\
 &=& (p\oplus q)_1'= p_1 + e^{-\ell p_0 }q_1 + \xi_1\big(\frac{1 - e^{-2\ell (p_0 + q_0)} }{2 \ell}+\\ 
 &-&\ell (p_1 + e^{-\ell p_0 } q_1)^2  + \frac{\ell}{2} |p + e^{-\ell p_0} q|^2 \big) \,,\nonumber
\end{eqnarray}
on the other hand the composition of the boosted momenta gives
\begin{eqnarray}
(p'\oplus  q')_1&=&p_1'+ e^{-\ell p_0' } q_1'= p_1 + e^{-\ell p_0}q_1 + \nonumber\\
&+&\xi_1\big(\frac{1 - e^{-2 \ell p_0} + e^{-\ell p_0 } -  e^{-\ell(p_0 + 2 q_0 )}}{2 \ell} +\\
&-&\ell (p_1^2+e^{-\ell p_0 } p_1 q_1+e^{-\ell p_0} q_1^2)+ \frac{\ell}{2}(|p|^2+ e^{-\ell p_0} |q|^2)\big)\,.\nonumber
\end{eqnarray}
Confronting those two results we can verify relation \eqref{neqback} just calculating
\begin{eqnarray}
(p\oplus q)_1'&-&(p'\oplus  q')_1=\nonumber\\
&=&\xi_1 \big(\frac{e^{-2 \ell(p_0 + q_0)}}{
 2 \ell} (e^{\ell p_0}-1 - e^{\ell(p_0 + 2 q_0)} \times\\
&\times& (1 +  \ell^2 (2 q_1^2- |q|^2 +2 (p_2 q_2+ p_3 q_3 )) ) +  e^{2\ell q_0 } (1 -  \ell^2(2  q_1^2 + |q|^2))\big)\,.\nonumber
\end{eqnarray}
At first sight this feature may suggest some kind of unreliability of the $\kappa$-Poincar\'{e} relativistic framework, since apparently two boosted observers could not have any way to come to an agreement about the result of some particles' interaction. However this is just a misleading interpretation based on the naive assumption that the generators should act trivially on plane waves as expected from a formalization based on simple lie algebras \eqref{trivialcopr}:
\begin{equation}
\mathcal{N}_i\vartriangleright (e^{i p_\mu x^\mu} e^{i q_\nu x^\nu})=(\mathcal{N}_i\vartriangleright e^{i p_\mu x^\mu}) e^{i q_\nu x^\nu} +  e^{i p_\mu x^\mu} (\mathcal{N}_i\vartriangleright e^{i q_\nu x^\nu})\,,
\end{equation}
while, according to the nontrivial boost coproduct \eqref{boostcopr}, the boost action on coupled plane-waves should be
\begin{eqnarray}
\mathcal{N}_i\vartriangleright (e^{i p_\mu x^\mu} e^{i q_\nu x^\nu})&=&(\mathcal{N}_i\vartriangleright e^{i p_\mu x^\mu}) e^{i q_\nu x^\nu} +\\  &+& (e^{-\ell P_0}\vartriangleright e^{i p_\mu x^\mu}) (\mathcal{N}_i\vartriangleright e^{i q_\nu x^\nu}) +\nonumber\\
&-&i \ell\epsilon_{ijk}(P_j\vartriangleright e^{i p_\mu x^\mu}) ({\cal R}_k\vartriangleright e^{i q_\nu x^\nu})\,.\nonumber\label{nontrivactboost}
\end{eqnarray}
The action at \eqref{nontrivactboost} should seem not easily manageable, however in \cite{GiuFla,GiuPalm} ({\it see} also \cite{MicGiu2022} for a clearer explanation) was suggested to re-express such feature as some sort of \emph{backreaction} of momenta on the boost parameter $\xi\vartriangleleft (p,q)$, as already formalized in \cite{Majidback}. The vertex conservation in $\kappa$-Poincar\'{e} is in general satisfied by
\begin{equation}
k'_\alpha(\xi)= (p'(\xi\vartriangleleft q)\oplus q'(\xi\vartriangleleft p))_\alpha\,.\label{backgenerica}
\end{equation}
In the bicrossproduct basis, however, at first order in the rapidity parameter $\xi$ equation \eqref{backgenerica} reduces to:
\begin{equation}
k'_\alpha(\xi) \simeq (p'(\xi)\oplus q'(\xi e^{-\ell p_0}))_\alpha\,.
\end{equation}
If one was interested to study such feature in the ultrarelativistic regime, a generalization of this last equation working well beyond the first order is provided in \cite{GiuFla,Majidback}, which is:
\begin{equation}
\xi\vartriangleleft p_i=2\arcsin\left(\frac{e^{-\ell p_0}\sinh\frac{\xi}{2}}{\sqrt{(\cosh\frac{\xi}{2}-\ell p_i\sinh\frac{\xi}{2})^2 - (e^{-\ell p_0}\sinh\frac{\xi}{2})^2}}\right)\,.\nonumber\label{MajidBackreaction}
\end{equation}
However this formalization gives a coherent description of particles' vertices under boost in $\kappa$-Poincar\'{e}, the problem related to the dependence of the reaction result to particles' ordering, already present in the non-boosted vertex, becomes even more problematic: if $p\oplus q \neq q\oplus p$, and therefore $p'(\xi)\oplus q'(\xi\vartriangleleft p) \neq q'(\xi)\oplus p(\xi\vartriangleleft q)$, when measuring $k$ and $k'$ should different observers agree on which particle comes first in the momenta composition? How should nature know about the particles momenta ordering that we are writing on paper? 
In the quantum-gravity framework this problem have been addressed on many different angles \cite{PapocchioSmolinFreidel,Trevisan,GiuLeo,LeoVertici}, and still is an open issue, on the other hand in a classical curved-spacetime framework such ordering mechanisms should not imply the same kind of issues, since there are clear rules governing the transformations between different observers in General Relativity.

\section{Backreaction in deSitter}

\begin{figure}[h!]
	\centering
	\framebox{\includegraphics[scale=0.35]{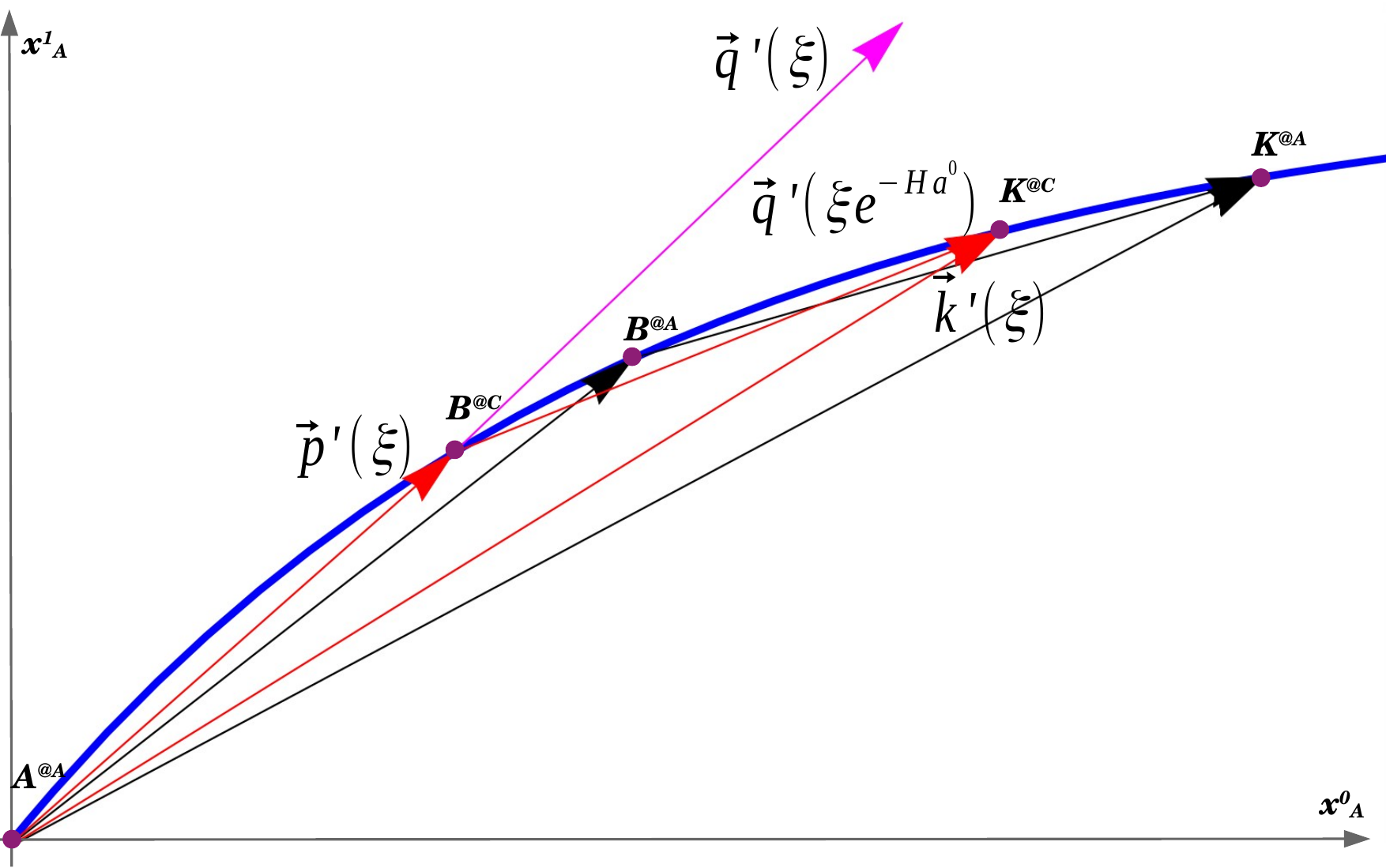}}
	\caption{\footnotesize The vector sum $\vec{k}=\vec{p}\oplus \vec{q}$ as expressed in the frame at rest (in black) and in the boosted one (in red). The Magenta vector represents $q'$ without taking into account the backreaction on the rapidity parameter $\xi$. The values fixed for the parameters are $H=1$, $a^0=b^0=1$ and $\xi=0.5$.}
	\label{fig:TriangoloBoost}
\end{figure}

In Section \ref{sec:OplusdSspacetime} we have defined three vectors $\vec{p},\vec{q},\vec{k}$ as expressed in Alice's ($@A$) and Bob's ($@B$) coordinate sets:
\begin{equation}
\begin{array}{c}
p_{@A}=(a^0,a^1)\, , \, q_{@B}=(b^0,b^1)\,,\\
q_{@A}=(b^0,b^1 e^{-H_\Lambda a^0}) \,,\, k_{@A}=(a^0+b^0,a^1+b^1 e^{-H_\Lambda a^0})\,.
\end{array}
\end{equation}
We also defined a composition law $\oplus$ so that those three vectors are forced to close a vector sum along a curved de Sitter worldline (see Figure \ref{fig:ABdeSitter}):
\begin{equation}
\vec{k}_{@A}=\vec{p}_{@A}\oplus \vec{q}_{@B}\,.
\end{equation}
We will define the observer Camilla as the boosted observer along the direction $1$, sharing it's origin with Alice. The vectors' components will transform under boost according to \eqref{dSspacetimetrasf}. Since a boost is well defined only in the origin we can directly boost only the vectors whose origin is in Alice's and Camilla's origin
\begin{equation}
p_{@C}=(p_{@A})'(\xi)\;,\;k_{@C}=(k_{@A})'(\xi)
\end{equation}
Boosting directly Bob's coordinates won't allow-us to close the vector sum in Camilla's frame, as clearly represented by the Magenta arrow in Figure \ref{fig:TriangoloBoost}.

Therefore for what concerns $\vec{q}$, whose origin lies in Bob, we will have to take into account also an effect on the rapidity, then
\begin{equation}
q_{@D}=(q_{@B})'(\xi\vartriangleleft p_{@A})\,,
\end{equation}
where given our choice of coordinates the action $\xi\vartriangleleft p_{@A}$ will be defined as the spacetime dual of equation \eqref{MajidBackreaction}. Vector $q_{@D}$ won't be the expression of $q_{@B}$ in Camilla's system, but in a translated one (say Diego). therefore if we want to compose $p$ and $q$ we need to remind that in curved spacetime we will have
\begin{eqnarray}
(k_{@A})'(\xi)&=&k_{@C}=p_{@C}+q_{@C}=p_{@C}\oplus q_{@D}=\nonumber\\
&=&(p_{@A})'(\xi)\oplus (q_{@B})'(\xi\vartriangleleft p_{@A})\,.\label{eq:BoostedTriangle}
\end{eqnarray}
Let's now try to perform this calculation one step at the time. For sake of simplicity we will limit to first order in $\xi$, which in a cosmological framework (non-ultrarelativistic regime) is perfectly justified. We will also assume the boost to be aligned with the signal direction ($x^2,x^3=0$). Therefore:
\begin{eqnarray}
(k_{@A})'(\xi)&\simeq& k_{@A}-i\xi[\mathcal{N}_1,k_{@A}]=\\
&=&\left\{
\begin{array}{l}
a^0 + b^0 - \xi (a^1 + b^1 e^{- H_{\Lambda}a^0})  \\
a^1+b^1 e^{-H_{\Lambda} a^0} + \xi \big(\frac{e^{-2 H_{\Lambda} (a^0+b^0)} - 1}{2 H_{\Lambda}} +\frac{H_{\Lambda}}{2} (a^1 + b^1 e^{- H_{\Lambda} a^0})^2 \big)
\end{array}\right.\,,\nonumber\\
\nonumber\\
(p_{@A})'(\xi)&\simeq& p_{@A}-i\xi[\mathcal{N}_1,p_{@A}]=\\
&=&\left\{
\begin{array}{l}
a^0-\xi a^1  \\
a^1 - \xi\left(\frac{1 - e^{-2 H_{\Lambda} a^0}}{2 H_{\Lambda}} - \frac{H_{\Lambda}}{2} (a^1)^2\right)
\end{array}\right.\,,\nonumber\\
\nonumber\\
(q_{@B})'(\xi&\vartriangleleft& p_{@A})\simeq q_{@B}-i\xi e^{-H_{\Lambda} a^0} [\mathcal{N}_1,p_{@A}]=\\
&=&\left\{
\begin{array}{l}
b^0-\xi e^{-H_{\Lambda} a^0} b^1  \\
b^1 - \xi e^{-H_{\Lambda} a^0}\left(\frac{1 - e^{-2 H_{\Lambda} b^0}}{2 H_{\Lambda}} - \frac{H_{\Lambda}}{2} (b^1)^2\right)
\end{array}\right.\,.\nonumber
\end{eqnarray}
Gathering all together, and taking into account that we lie along a deSitter worldline $x^1=(1-e^{-H_{\Lambda} x^0})/H_{\Lambda}$, it is straightforward to verify equation \eqref{eq:BoostedTriangle}.\\
Then, while in the $\kappa$-Poincar\'{e} approach to quantum gravity a clear physical reason for the ordering dependence of momenta vertices is still debated, in the deSitter framework the interpretation is pretty easy. The backreaction in curved spacetime formalizes the need to "close" the vector sum between vectors laying on different observers frames ({\it see} Fig.\ref{fig:TriangoloBoost}). The ordering is relative to the observer who wants to analyze and confront signals originating from different translated sources, while taking into account apparent velocity discrepancies.

\section{In summarising de Sitter Hopf algebra}\label{sec:Summarizing}

It is common knowledge that $\kappa$-Poincar\'{e} is a deSitter-inspired curved momentum-space Hopf algebra. What is not-so-common knowledge, however, is that also in deSitter framework we can recognize an Hopf-algebra structure. In fact, in summarizing what we obtained so far, the deSitter deformed Lorentz generators $\mathcal{N}_i ,\mathcal{R}_j$ together with coordinates $x^\mu$ close an Hopf algebra with algebraic sector
\begin{eqnarray}
&[x^\mu,x^\nu]=0 \;,\;\;\;&\\ 
&[\mathcal{N}_i , x^0]=-i\delta_{ij}x^j\;,\;
 [\mathcal{N}_i , x^j]=-i\delta_i^j\left(\frac{1-e^{-2 H_\Lambda x^0}}{2H_\Lambda}+\frac{H_\Lambda}{2}|x|^2\right) +iH_\Lambda x^k x^j \delta_k^i \; ,&\nonumber\\
&[\mathcal{R}_i ,x^0]=0 \; , \;\;\;[\mathcal{R}_i ,x^j]=-i\epsilon_{ilk}\delta^{lj} x^k\, ,&\nonumber\\ 
&[\mathcal{R}_i , \mathcal{R}_j]=-i\epsilon_{ijl}\delta^{lk}\mathcal{R}_k\;,\;[\mathcal{N}_i ,\mathcal{N}_j]=i\epsilon_{ijl}\delta^{lk}\mathcal{R}_k\;,\;[\mathcal{R}_i ,\mathcal{N}_j]=-i\epsilon_{ijl}\delta^{lk}\mathcal{N}_k\,,&\nonumber
\end{eqnarray}
and whose coalgebraic sector is dual (or in more appropriate words completely isomorphic) to the $\kappa$-Poincar\'{e} one:
\begin{eqnarray}
&\Delta x^0 = x^0\otimes \mathbb{1} + \mathbb{1}\otimes x^0\,,\nonumber&\\
&\Delta x^i = x^i\otimes \mathbb{1} + e^{-H_\Lambda x^0} \otimes x^i\,,\nonumber&\\
&{\cal S}(x^0)=-x^0\,,\;\;\;{\cal S}(x^i)=-e^{H_\Lambda x^0}x^i\,,&\\
&\Delta {\cal N}_{i} ={\cal N}_{i}\otimes \mathbb{1} + e^{-H_\Lambda x^0}\otimes {\cal N}_{i}-iH_\Lambda \epsilon_{mjk}\delta^{mi} x^j\otimes {\cal R}_{l}\delta^{lk}\,,\label{dScopr}&\nonumber\\
&{\cal S}({\cal N}_{i})=-e^{H_\Lambda x^0}{\cal N}_{i}+i H_\Lambda \epsilon_{mjk} x^j \delta^{mi}\delta^{lk}{\cal R}_{k}\,,\nonumber&
\end{eqnarray}
Such an Hopf algebra defines also an algebra between the different observers frames, so that
\begin{equation}
\Delta x^\mu(x_{@A},x_{@B})=(x_{@A}\oplus x_{@B})^\mu\,,\label{coprsum}
\end{equation}
which is associative given the coassociativity axiom of Hopf algebras:
\begin{equation}
((x_{@A}\oplus x_{@B})\oplus x_{@C})^\mu=(x_{@A}\oplus (x_{@B}\oplus x_{@C}))^\mu\,. \label{eq:coassociativity}
\end{equation}
As well as the coproduct defines the composition law, on the other hand the antipode defines the inversion $\ominus$:
\begin{equation}
{\cal S}(x^\mu)_{@A}=\ominus x^\mu_{@A}\,.
\end{equation}
The last element we need to define is the origin of the observer's framework which in turn can be defined through the deSitter Hopf algebra counity :
\begin{equation}
x^\mu(\underline{0})=\epsilon(x^\mu)=((\ominus x_{@A})\oplus x_{@A})^\mu\,.
\end{equation}
The spacetime observers' algebra is then equipped with a neutral object $\underline{0}$, a composition rule $\oplus$ and the inversion $\ominus$, for which the associativity property holds. Composition law and inversion can be intuitively described as sliding along the worldline connecting different observers. Of course the worldlines' shape depends on the initial choice of coordinates \eqref{dSflatcoordinates} from which in turn we can define a different Killing equation, different symmetry generators and of course a different algebra. A different set of coordinates in the observers' framework will then imply a change in the basis of the Hopf algebra we will rely on, which in turn will bring to a different definition of the composition law $\oplus$ and its inversion $\ominus$, just as expected, since also the worldlines would change form accordingly. This means that there is a well defined relation between spacetime diffeomorphism and the choice of the Hopf algebra basis.\\
This idea of sliding along the worldlines helps-us to develop an intuition also for what concerns the boost and the backreaction: while in flat spacetime one can imagine that a boosted observer will perceive all the other observers to be boosted relatively to him with the same parameter $\xi$ (since in special relativity each observer will transform independently from the others), in curved spacetime this won't be true anymore and the boost performed in some observer's origin will act differently on the others, according to their distance along the worldline, so that if one observer at rest expresses some point position as the composition of a few vectors
\begin{equation}
r_{@A}=p_{@A}\oplus q_{@B}\oplus k_{@C}\,,
\end{equation}
a boosted observer $D$ in the origin of $A$ will have:
\begin{eqnarray}
r_{@D}&=&r_{@A}'(\xi)=p_{@A}'(\xi)\oplus q_{@B}'(\xi\vartriangleleft p_{@A})\oplus
 k_{@C}'(\xi\vartriangleleft p_{@A}\vartriangleleft q_{@B})=\\
&=&p_{@A}'(\xi)\oplus q_{@B}'(\xi\vartriangleleft p_{@A})\oplus k_{@C}'(\xi\vartriangleleft (p_{@A}\oplus q_{@B}))\,.\nonumber
\end{eqnarray} 
Of course the identification 
\begin{equation}
\xi\vartriangleleft p_{@A}\vartriangleleft q_{@B}=\xi\vartriangleleft (p_{@A}\oplus q_{@B})\,,
\end{equation}
is implied by the associativity of $\oplus$, but it also expresses that the right action on $\xi$ grows the more the boosted observer origin is far away from the origin of a signal along the signal's worldline. In our context this may imply that an observer will perceive closer objects as more boosted than further ones, which may have interesting implications from the phenomenological point of view.\\

\section{Nontrivial composition law and cosmological redshift}
\begin{figure}[h!]
	\includegraphics[scale=0.95]{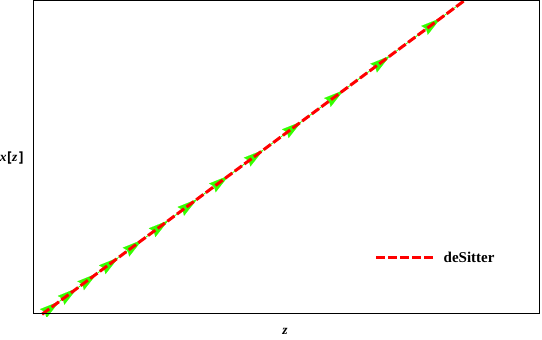}
	\includegraphics[scale=0.95]{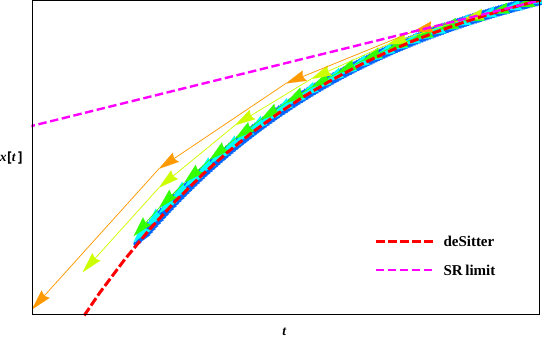}
	\caption{\footnotesize On top $n=13$ locally Special relativistic vectors in a $z$ vs. $x^i$ graphic are composed according to nontrivial coproducts \eqref{zxcopr}. Of course there is no need for a large precision, since a straight line is easy to reproduce, however an interesting feature to be noticed is that the vectors are wider and wider, due to the nontrivial composition laws of both variables. In the graphic below the reconstruction of the deSitter photons worldlines, with respect to the photon's time of emission ($-t$). The larger $n$, the more accurate is the reconstruction of the deSitter worldlines $x^i(t)$. The values of $n$ represented are (4,6,20,30,1000).}
	\label{fig:FreccetteZdS}
\end{figure}

The application of the previously introduced mathematical framework to the study of cosmological and astrophisical observations, would, of course, require further exploration of more complex non-maximally-symmetric spacetime models. Such investigation might need the development of {\it ad hoc} techniques (also inspired to QG literature {\it see} for instance \cite{Slicing}), in order to study the  Hopf-algebraic features of signals in a non-empty universe.\\
Since the aim of this paper is the introduction of the physical interpretation of some basic and general Hopf-algebraic features, it may be better to leave such a complex task to future investigations. However a simple way to introduce a more concrete element in our analysis is the study of redshift coproduct and its practical implications.\\
In cosmology the sygnals time-of-emission is an inference of the model. The actual observables are, on the other hand, distance (from the receiver point of view) and redshift $z$. The definition of redshift is the very well known $z=(\lambda_{observed}-\lambda_{emitted})/\lambda_{emitted}$, which in our simple model translates into
\begin{equation}
	z=e^{-H_\Lambda x^0}-1\,.\label{redshift}
\end{equation}
Since in Section \ref{sec:Composition1} we used spacetime translations to provide a physical interpretation to coproduct, it may make sense now to try a similar approach to obtain a redshift coordinate coproduct. Since using the definition \eqref{redshift} we have:
\begin{equation}
	b^\alpha\{\Pi_\alpha , z\}=-b^0 H_\Lambda e^{-H_\Lambda x^0}=-b^0 H_\Lambda(1+z)\, ,
\end{equation}
applying iteratively the procedure at eq. \eqref{GenAct} to Alice observer's redshift coordinate $z_a$, we obtain:
\begin{equation}
	b^\alpha\Pi_\alpha \vartriangleright z_a = z_a + z_b \, e^{-H_\Lambda a^0}\, ,
\end{equation} 
where of course $z_b=e^{-H_\Lambda b^0}-1$ and $a^0$ is the inferred Alice time coordinate in the observer framework. Therefore redshift coproduct can be defined as:
\begin{equation}
	\Delta z = z \otimes \mathbb{1} + e^{-H_\Lambda x^0} \otimes z\,.
\end{equation}
However since, as we already stated, time-coordinate (which is the only one that scales linearly) can be inferred only after the definition of a model, basing on it an analysis aimed to the study of cosmological signals, may lead to a fundamental inconsistency. Therefore it is useful to redefine coordinate composition (and their related coproducts) in terms of physical observables:
\begin{equation}
	z_a\oplus_z z_b = e^{-H_\Lambda x_a^0 - H_\Lambda x_b^0}-1 = z_a + z_b e^{-H_\Lambda x_a^0}= z_a + z_b(1+z_a)\,,\label{compz}
\end{equation} 
and of course for what concerns space coordinates:
\begin{equation}
	x_a^i\oplus_x x_b^i = x_a^i + x_b^i e^{-H_\Lambda x_a^0} = x_a^i +x_b^i\cdot(1+z_a)\,,\label{compx}
\end{equation}
which lead to:
\begin{equation}
	\begin{array}{c}
		\Delta z = z \otimes \mathbb{1} + (1+z) \otimes z\, ,\\
		\Delta x^i = x^i \otimes \mathbb{1} + (1+z) \otimes x^i\,.\label{zxcopr}
	\end{array}\, .
\end{equation}
As in Section \ref{sec:Composition1}, also \eqref{compz} and \eqref{compx} express vector coordinates composition along a worldline $x^i(z)$, the only difference with the $x^i(x^0)$ case, is that now both coordinates evolve nontrivially at the same way, which is to be expected, since photons deSitter worldlines in terms of $z$ are linear:
\begin{equation}
	x^i(z)=\int_{-x^0}^0 \frac{c}{a(t)}dt=\int_{0}^z \frac{c}{a(z)}\frac{dZ}{H_\Lambda (1+Z)}=\frac{c}{H_\Lambda}z\,.
\end{equation}
Calculations in deSitter are simple, but a few problem may arise when dealing with more complex models, in which the number of independent Killing fields is six instead of deSitter's ten. The way to deal with this minor issue is to imagine a series of translated locally flat observers, with flat translation parameters, $(t,x,z)\rightarrow(\epsilon^0,\epsilon^1=c\cdot\epsilon^0,\epsilon^z=(H_\Lambda/c)\epsilon^1)$, and express the point of view of the observer at the origin as the total nontrivial composition of $n$ locally flat frames coordinates. Any point $P(z,x^i)$ along a photon worldline can be then expressed as:
\begin{equation}
	P(z,x^i)=(\epsilon^z_A,\epsilon^i_A)\oplus_{z,x}(\epsilon^z_B,\epsilon^i_B)\oplus_{z,x}\dots\oplus_{z,x}(\epsilon^z_\Omega,\epsilon^i_\Omega)\,,\label{sommona}
\end{equation}
where, of course, the indices $\oplus_{z,x}$ below the composition sign, remind us that it stands for different rules for $z$ and $x^i$, according to \eqref{compz} and \eqref{compx}. In our case the result of this complex mechanics is however just a straight line. In order to apply this procedure to a little more interesting curve, we can just re-express the same worldline in terms of space and time, $x^i(x^0)$ \eqref{dSworldline}, using $n$ local SR vectors $(\epsilon^0,\epsilon^1=c\cdot \epsilon^0)$, composed according to \eqref{kcomposed}. Both those curve-reconstruction are shown in Figure \ref{fig:FreccetteZdS}.\\

As we can observe in Figure \ref{fig:FreccetteZdS}, the locality of the flatness requirement is of paramount importance since the composition of not-narrow-enough vectors fails to collapse to the correct deSitter worldline. However no complex numerical simulations are required in order to obtain satisfactory results: a few hundreds elements composition is practically Indistinguishable from the curve itself.\\
 Of course this first exploration of spacetime coproduct expressed in terms of redshift is effective only in a maximally-symmetric deSitter framework. The main issue, in order to apply an Hopf algebraic mathematical framework to a realistic physical context, is of course related to how spacetime symmetries breakdown in presence of energy and matter (bright or dark at will) should be addressed.

\section{Cosmological deSitter slicing}

As done in last section, from now on we will use the coproduct to reproduce the integration on a curved spacetime $x^1(x^0)=\int_0^{x^0} c\cdot\frac{dt}{a(t)}$ to reconstruct photons' path from source to observer. If we express with $\epsilon_{(n)}=(\epsilon^0,\epsilon^1)$ the locally special relativistic vector for which of course $\epsilon^1=c\cdot\epsilon^0$, the operator $\oplus$ imposes the right amount of correction in order to reconstruct the right deSitter trajectory from the emitter point of view, while an observer receiving the signal, in order to infer the emission spacetime coordinates should instead compose the vectors using the antipode $\ominus$:
\begin{eqnarray}
	\left(t,\int_0^t e^{-H_{\Lambda} t}dt\right)\simeq \vec{\epsilon}_I\oplus\vec{\epsilon}_{II}\oplus \ldots \oplus \vec{\epsilon}_{N}\,,\label{eq:PrimitiveSlicing}\\
	\left(-t,\int_{0}^{-t} e^{H_{\Lambda} t}dt\right)\simeq \vec{\epsilon}_I\ominus\vec{\epsilon}_{II}\ominus \ldots \ominus \vec{\epsilon}_{N}\,.
\end{eqnarray}
As expected for $\epsilon^0\rightarrow 0$ and $N\rightarrow \infty$ the sum converges to the integral, and the vectors align themselves to the deSitter cuve more and more precisely, as shown in Fig.\ref{fig:FreccetteZdS}. In this section, in order to show the graphics in terms of Megaparsec/year, the speed of light and the cosmological constants will be expressed in units of: $c=3.066\cdot 10^{-7}\,\text{Mpc}/\text{y}$, $H_0=7.568\cdot 10^{-11}\, 1/\text{y}$ and the time units are expressed in fractions ($1/n$) of the extimated life of the universe $t_u=13.8 \cdot 10^9$ years.\\

Of course the possibility to describe a particle's trajectory using the right composition of infinitesimal vectors was pretty much granted, however it is not as well granted how to extend the Hopf algebraic formalism cases in which spacetime symmetries are broken. Let's for instance recover the first Friedman equation \eqref{modFried1}, in a mostly flat ($k=0$) deSitter universe with a little of homogeneously distributed matter with the same amount of particles from the beginning $\rho = \rho_0/a(t)^3$:
\begin{equation}
	\left(\frac{\dot{a}}{a}\right)^2 = \chi \frac{\rho_0}{a^3} + H_\Lambda^2\,,
\end{equation}
or, defining $\chi \rho_0 = H_0^2\Omega_{m 0}$ and $H_\Lambda^2=H_0^2\Omega_{\Lambda 0}$:
\begin{eqnarray}
	\dot{a}^2 = H_0^2\Omega_{m 0} a^{-1} + H_0^2\Omega_{\Lambda 0} a^2\,.
\end{eqnarray}

\begin{figure}[h!]
	\centering
	\includegraphics[scale=1.2]{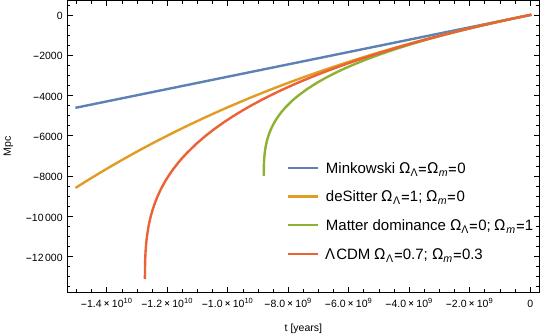}
	\caption{\footnotesize An observer in its origin receives photon worldlines emitted in a distant past $-t$ for different cosmological models.}
	\label{fig:BBAlice}
\end{figure}

\begin{figure}[h!]
	\centering
	\includegraphics[scale=1.2]{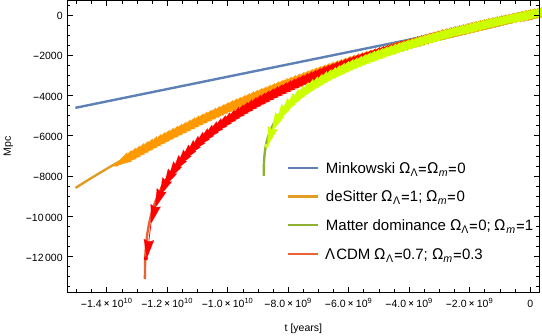}
	\caption{\footnotesize Phenomenological coproduct deformed composition of Special Relativistic vectors $(\epsilon^0,\epsilon^1)$, along the worldlines shown in Fig.\ref{fig:BBAlice}. The observer in $(0,0)$ reconstructs the photon paths as an exchange of signals between $n=100$ intermediate observers.}
	\label{fig:PhenCoproduct}
\end{figure}

The solution of this equation is
\begin{equation}
	a(t) = \sqrt[3]{\frac{\Omega_{m 0}}{\Omega_{\Lambda 0}}} \sinh^{\frac{2}{3}}\left(\frac{3}{2}H_0\sqrt{\Omega_{\Lambda 0}}t + \text{arcsinh} \sqrt{\frac{\Omega_{\Lambda 0}}{\Omega_{m 0}}} \right)\,,
\end{equation}
which in case respectively of an empty universe ($\Omega_{m 0}\rightarrow 0$) or no cosmological constant ($\Omega_{\Lambda 0}\rightarrow 0$) would reduce to
\begin{equation}
	a(t)\rightarrow\left\{\begin{array}{lr}
		\Omega_{m 0}\rightarrow 0 & e^{H_0\sqrt{\Omega_{\Lambda 0}}t}=e^{H_\Lambda t}\\
		\Omega_{\Lambda 0}\rightarrow 0 & \left(1+\frac{3}{2}H_0\sqrt{\Omega_{m 0}}t\right)^{\frac{2}{3}}
	\end{array}
	\right.\,.
\end{equation}
The Hubble constant is defined as $H(t)=\dot{a}/a$, then:
\begin{equation}
	H(t)=H_0\sqrt{\Omega_{\Lambda 0}}\coth\left(\frac{3}{2}H_0\sqrt{\Omega_{\Lambda 0}}t + \text{arcsinh} \sqrt{\frac{\Omega_{\Lambda 0}}{\Omega_{m 0}}} \right)
\end{equation}
which again in the limits $\Omega_{m 0}\rightarrow 0\,,\Omega_{\Lambda 0}\rightarrow 0$ evolves in time as:
\begin{equation}
	H(t)\rightarrow\left\{\begin{array}{lr}
		\Omega_{m 0}\rightarrow 0 & H_0\sqrt{\Omega_{\Lambda 0}} = H_\Lambda\\
		\Omega_{\Lambda 0}\rightarrow 0 & \frac{H_0\sqrt{\Omega_{m 0}}}{1+\frac{3}{2}H_0\sqrt{\Omega_{m 0}}t}\label{eq:Hlimits}
	\end{array}
	\right.\,.
\end{equation}
By specifying the values of $\Omega_{m 0}\rightarrow (1,0,0.3)$ and $\Omega_{\Lambda 0}\rightarrow (0,1,0.7)$, one obtains respectively the matter dominance, deSitter and $\Lambda\text{CDM}$ cosmological models. The reproduction of the photons' worldlines received by the observers for any value of the $\Omega$ parameters are depicted in Figure \ref{fig:BBAlice}.\\
It is well known that the introduction of matter breaks spacetime symmetries, for instance symmetry under spatial rotations is kept while broken under boosts. How should we keep the whole Hopf algebraic structure for the matter-dominated universe and $\Lambda\text{CDM}$ models? One naive solution may be to imagine some sort of "phenomenological" coproduct
\begin{equation}
	\Delta x^i = x^i \otimes \mathbb{1} + a^{-1}(x^0)\otimes x^i \,,\label{eq:PhenoCoproduct}
\end{equation}
which may not be completely justified from the mathematical point of view, but that physically speaking gets the job done, reproducing the models' photons worldlines from source to observer. Using this approach, in fact equation \eqref{eq:PrimitiveSlicing} would become
\begin{equation}
	x_{@A}^1\oplus x_{@B}^1\oplus x_{@C}^1\oplus\,_{\dots} =x_{@A}^1\oplus x_{@B}^1 a^{-1}(x^0_{@B}) \oplus x_{@C}^1 a^{-1}(x^0_{@C}) \oplus\,_{\dots}\, .
\end{equation}
Applying this logic a cospicuos number of times the curves obtained integrating $w(t)=\int_{-t}^0 \frac{c}{a(x^0)}dx^0$ can be numerically replicated through this phenomenologically-deformed vector composition ({\it see} Fig.\ref{fig:PhenCoproduct}).\\
This formalization, however useful from the physical point of view, doesn't help us finding a way to apply Hopf algebras to the study of cosmology, without which this analysis would be a mere theoretical speculation on the Hopf-algebraic aspects of deSitter in-vacuo solution of FLRW equations.\\
The possibility to unravel our issue can be provided again by quantum gravity literature where such a problem was already dealt within the DSR framework.

\subsection{The slicing approach}

\begin{figure}[h!]
	\includegraphics[scale=0.95]{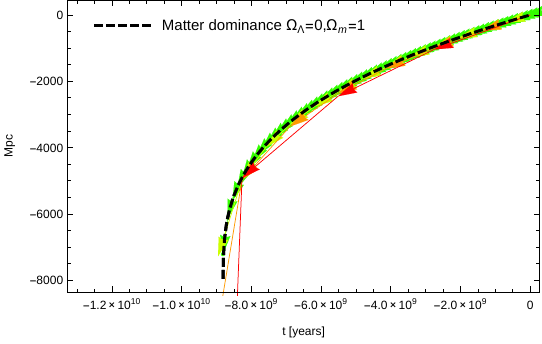}
	\includegraphics[scale=0.95]{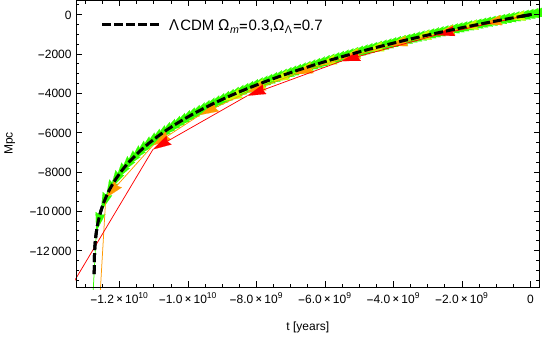}
	\caption{\footnotesize The slicing approach with matter dominance (on the left) and $\Lambda\text{CDM}$ (on the right). The red, orange, light green and green arrows represent special relativistic vectors, composed with deformed composition law within $n=5,10,20,100$ slices of locally deSitter frameworks. The dotted black lines are the worldlines solutions obtained by numerically integrating $c\cdot a^{-1}(t)$.}
	\label{fig:GraficoSlicing}
\end{figure}

In \cite{Slicing}, in order to study the interplay between Planck-scale effects and spacetime curvature and in particular the effects of Planck-scale modifications of the dispersion relation on photon propagation along cosmological distances, the metric of FLRW  in a matter dominated universe was sliced into locally deSitter spacetime portions. In this approach, the element connecting a matter dominance worldline with each of its $N$ deSitter slices is the running of the cosmological constant $H(t)$, according to \eqref{eq:Hlimits} when $\Omega_{\Lambda 0}=0$ and $\Omega_{m 0}=1$, and the composition of photon trajectories in the n-th slice, are described by $\text{Bob}_N$ to be
\begin{equation}
	x^{B_N}(t^{B_N})_n = x_{O_A}^{B_N}+\sum_{k=1}^{n-1} e^{\sum_{s=k+1}^N H_s \epsilon^0} \cdot \frac{e^{H_k \epsilon^0}-1}{H_k}\,.
\end{equation}
In our case we will adopt a similar methodology: we will split the photon trajectory into $N$ intermediate observers whose origin are connected one-by-one by special relativistic vectors from the emitter point of view, where the deSitter deformation is due to the coproduct. Every observer will live in a locally deSitter slice with different cosmological constant $H_{@I}$ (where the $I$ index denotes the observer). This will locally keep unbroken deSitter Hopf algebra, the only real issue is that we would loose the overall coassociativity property \eqref{eq:coassociativity} of the observers' coproducts. Therefore the composition law, taking into account the cosmological constant running, becomes:
\begin{eqnarray}
	\int_0^t\frac{c}{a(x^0)}dx^0&\simeq& x^1_{@A} + x^1_{@B} e^{-H_{@B}\epsilon^0} + x^1_{@C} e^{-H_{@B}\epsilon^0}e^{-H_{@B}\epsilon^0} + \dots =\nonumber\\ 
	 &=& x^1_{@A} + \sum_{I=1}^N x^1_{@I} e^{-\sum_i^I H_i \epsilon^0}
\end{eqnarray}
As already mentioned, in \cite{Slicing} the slicing approach was applied on a cosmological model with the cosmological constant running in accordance with the relation $H(t)=H_0/(1+\frac{3}{2}H_0 t)$. In order to be more confident on the generality of this procedure, we will apply it both on matter dominance and $\Lambda\text{CDM}$ models. We will go backward in time along a worldline, splitting time in units of $\epsilon^0 = -t_u/N$. Every observer's origin will be localized in terms of Bob's coordinates at
\begin{equation}
	x^1_{B_{n+1}}(x^0_{@B_{n+1}}=0)=x^1_{B_n}(x^0_{@B_{n}}=0)-\frac{t_u}{N}\cdot e^{-\sum_{i}^{n} (H_n t_u/N)}\,,
\end{equation}
where being each vector locally special relativistic $\epsilon^1 = c\cdot |\epsilon^0|$ and where every $H_n$ should be calculated according to the models we are taking into account.\\
As shown in Figure \ref{fig:GraficoSlicing} the bigger the number $N$ of deSitter slices, the better the vector composition will converge to the models' curves.\\

In this section we were able to reproduce signal propagation according to known cosmological models. With a little more effort this approach could also be used to reconstruct the photon worldlines without the need to assume any theoretical model, knowing only the evolution of the cosmological constant through the eras of the universe. However this would need a careful analysis on cosmological redshift and Hopf algebras which we leave to further explorations.

\section{Conclusions}
\noindent
In summarizing, we have seen that an Hopf algebraic aspects of deSitter model, although never acknowledged in the literature, can be found in many commonly and not-so-commonly studied cosmological frameworks.\\
Coproduct, for instance, is very naturally embodied in deSitter model as signal propagation in curved spacetime, and can be seen as a different way to reconstruct the already known photon worldlines shapes under different hypotheses about the configuration of the universe. Other Hopf algebraic features like backreaction are way less obvious to recognize in the general framework of deSitter cosmology, however, once correctly interpreted, may provide a different angle to analyze relativistic properties of curved spacetimes.\\
If on the one hand coproduct for coordinates may be seen just as a different way to sum or integrate portions of spacetime, the study of the deSitter "coalgebraic sector" may hold a few surprises when studying boost composition or the interaction between boosts and translations. In particular a more in-depth study of boost coproduct and in general the action of deSitter boost on different observers, subject to the relativistic relations summarized in Section \ref{sec:Summarizing}, may not be insignificant, in a field of study in which reconstructing objects redshift plays a crucial role.\\
Of course this article is intended just as a previous exploration regarding the application of Quantum Groups {\it $\grave{a}$ la Majid} for the study of General Relativity. Such a mathematical framework provides a wide set of tools to effectively dissect relativistic phenomena. If coproduct and backreaction are that interesting, what can we say about Weyl maps, star product or Drinfeld twists? We know what the cosmological constant $H_\Lambda$ is for cosmology, but what may be its interpretation in a Quantum Algebra framework?\\
In parallel with the mathematical analysis a further evolution of this work may be an intense phenomenological exploration, extending this approach to non-maximally-symmetric models of the universe (such as $\Lambda \text{CDM}$) in order to apply the mathematical model to real, actual cosmological data, studying wether this novel tools are able to suggest new features or new ways to analyze the old ones.\\
There is still a lot of work to be done in order to understand if this kind of in-depth-study may actually be productive from a physical perspective or not. However what I would like to communicate to all those working on Quantum Gravity and DSR is that there is some digging to do and that you have the right shovels. 

\section*{Acknowledgments}
{\it I would like to thank Antonino Marcian\'{o} for his suggestions and insightful conversations.\\
I acknowledge support from the Huzhou Rihua Information Technology Co., Ltd and Zhejiang Kelin Talent Development Co., Ltd in whose offices this paper was partially written.\\
I also thank an unknown (allegedly Russian) hacker for giving me the necessary time and peace of mind to set up this work, by devastating the computer system of my previous workplace.}\\

\bibliographystyle{apsrev4-2}
\bibliography{UniverseNotLie}

\end{document}